\documentclass[12pt]{article}
\usepackage[centertags]{amsmath}
\makeindex
\newcommand{\beq}{\begin{equation}}
\newcommand{\eeq}{\end{equation}}
\newcommand{\beqr}{\begin{eqnarray}}
\newcommand{\eeqr}{\end{eqnarray}}
\def\lsim{\raise0.3ex\hbox{$\;<$\kern-0.75em\raise-1.1ex\hbox{$\sim\;$}}}
\def\gsim{\raise0.3ex\hbox{$\;>$\kern-0.75em\raise-1.1ex\hbox{$\sim\;$}}}

\begin{document}
\begin{center}
{\large \bf   Generalized Set of  Boussinesq equations for surf zone region}

{ R. Dutta \footnote{Center of Higher Learning,
    University of Southern Mississippi, Bldg 1103, Stennis Space
    Center, MS 39529. {\it Email :}
    rdutt@orca.st.usm.edu(corresponding author)}}
and {J. Veeramony \footnote{GeoResources
    Institute, Mississippi State University, Bldg 1103, Stennis Space
    Center, MS 39529. {\it Email :} veeramon@gri.msstate.edu }}

\end{center}
\medskip

\begin {abstract}

In this report, generalized wave breaking equations are developed
using three dimensional fully nonlinear extended Boussinesq equations
to encompass rotational dynamics in wave breaking zone. The
derivation for vorticity distributions are developed from Reynold
based stress equations.

\end{abstract}


\section{Introduction}

Wave breaking is one of the most complex phenomena that occurs in the
nearshore region. During propagation of wave from deep to shallow
water, the wave field is transformed due to shoaling and close to the
shoreline, they become unstable and break. In the process of breaking,
energy is redistributed from fairly organized wave motion to small
scale turbulence, large scale currents and waves.

It has been shown by numerous researchers that Boussinesq-type
equations for varying water depth can describe nonlinear
transformation in the shoaling region quite well. In the last couple
of decades, a lot of research effort has gone into improving the
predictive capability of these equations in the intermediate
water-depth and close to the surf zone (see e.g. Nwogu [1983],
Madsen\cite{zou} , Wei [1995]). It was established that to extend the
validity of these equations to the deep water, higher order dispersive
terms will have to be retained, and to improve the predictive
capability close to breaking, the nonlinear terms will all have to be
retained. However, to model wave breaking, these models
use additional terms that artificially added to the momentum equation,
which would then reproduce the main characteristic of a breaking wave,
i.e. the reduction in wave height. For example, wave breaking in
FUNWAVE (FUNWAVE is based on the model described by Nwogu [1993])
is modeled by introducing momentum mixing term developed by Kennedy et
al [1999].

Most progress have been done for potential flow, starting with the
work of Nwogu [1993] and Madsen [1983]. Some work have also been done
to address partially rotational flows by Shen [2000]. In the breaking
region and in the surf zone, the wave breaking introduces vorticity
into the fluid. To address this problem, Veeramony \& Svendsen [2000] 
derived breaking terms
in Boussinesq equation assuming flow as two-dimensional rotational
flow. Here, the breaking process is modeled by assuming that vorticity
is generated in the roller region of the breaking wave and solving
vorticity transport equation to determine the distribution of the
vorticity. This naturally introduces additional terms in the momentum
equation which causes wave height reduction as well as changes in the
velocity field. However, since
this model is based on stream function formulation, it cannot be
trivially extended to three-dimensional flow. The phenomena of wave
breaking in Boussinesq equations are being modeled using quite few
techniques which can preserve the wave shape as well as include energy
dissipation mechanism.  Shen [2000] developed a generalized form
of Bousinesq equation in 3D vortical flow field with arbitrary
vorticity distribution up to O($\mu^{2})$. But he did not describe
momentum transport equation with full description of rotational flow.
Recently, Zou et al [2004] addressed the problem by including the
higher order terms in Boussinesq equation in 2D flow. This model
solves to vorticity distribution based on the parametric form taken
form surface roller data.  In this paper, we try to develop a general
form for breaking term for fully nonlinear set of Boussinesq equations
for three dimensional vortical flow field near surf zone region.
Derivation of breaking term from Reynold stress based vorticity
transport equation was also developed to describe rotational field as
a complete model of Veeramony [2000].

The paper is organized as follows: Section 2 discusses the
basic governing equations for continuity and momentum with
boundary conditions. Section 3 describes the equation for horizontal
and vertical velocity distribution for potential and rotational
components.  In section 4, the breaking term is derived for velocity
transport equation for fully nonlinear case and solved vorticity
transport equation analytically from fourier series expansion. In last
section, results were discussed with conclusion.

\section  {Basic Equations}

We consider a three-dimensional wave field with free surface
$\eta(x,y,t)$ propagating over a variable water depth $h(x,y)$.  As we
are primarily concerned with wave breaking, we only consider here wave
propagation in shallow water. Wave in this region can be characterized
by two non-dimensional parameters $\delta = {a}/{h}$ and $\mu =
{h}/{l} $ where $a$ is the characteristic wave amplitude and $l$ the
characteristic wave length.  The parameter $\mu$ is a measure of
frequency dispersion and $\delta$ that of the nonlinearity of the
wave. In this study, since we are only considering shallow water
waves, we only have to consider weakly dispersive waves (upto $
O(\mu^{ 2})$) but have to retain all nonlinear terms.

In this paper, the variables are non-dimensionalized using following
scaling:
\begin{subequations}
\begin{align}
 x &= \hat x /l, \; y = \hat y/l, \; z = \hat z /h, \; t =
 \hat t\sqrt{gh}/l,\\
\hat u &= \left(\delta\sqrt{gh}\right)u, \: 
\hat v = \left(\delta\sqrt{gh}\right)v, \; \hat w = \left(\delta
\mu \sqrt{gh}\right) w
\end{align}
\end{subequations}
where the $\;\hat{}\; $ represents the dimensional variables, $g$ is the
acceleration due to gravity, $u$ and $v$ are the horizontal components
of the velocity in the $x$ and $y$ directions respectively, $w$ is the
vertical velocity. We start with the Eulerian equations of continuity
and momentum in  nondimensionalized form for velocity field $ {\bf u} = (u,v,w)$ 
as:
\begin{align}
&\frac{\partial u} {\partial t} + {\delta} u\frac{\partial u}{\partial
  x} + {\delta}v\frac{\partial u}{\partial y} + {\delta}
w\frac{\partial u}{\partial z} + \frac{\partial p}{\partial x} = 0\\
&\frac{\partial v} {\partial t} + {\delta}v\frac{\partial v}{\partial
x} + {\delta} v\frac{\partial v}{\partial y} + {\delta}
w\frac{\partial v}{\partial z} + \frac{\partial p}{\partial y} = 0\\
{\delta \mu^{ 2}} &\frac{\partial w} {\partial t} + {\delta^{ 2}
\mu^{ 2}} u\frac{\partial w}{\partial x} + {\delta^{ 2}\mu^{
2}} v\frac{\partial w}{\partial y} + {\delta^{ 2} \mu^{ 2}}
w\frac{\partial w}{\partial z} + {\delta} \frac{\partial p}{\partial
z} + 1 = 0
\end{align}

Since the fluid flow is rotational, we also have three dimensional
 vorticity field $ {\bf s} = (s_x,s_y,s_z) $ in the
fluid defined as
\begin{equation}
 {\bf \bigtriangledown} \times  {\bf u } ={\bf s}  
\end{equation}
where $ \bigtriangledown = (\partial/\partial x, \partial/\partial y,
\partial/\partial z). $ 
The continuity equation then becomes, 
\begin{equation}
\bigtriangledown \cdot u + \frac{\partial w}{\partial z} = 0 
\end{equation}
Here $ \bigtriangledown \cdot u = (\partial u/\partial x,\partial v/\partial y). $
The above equations satisfy two boundary conditions for velocity at
bottom and at free surface.  At the free surface $ {\it z}=
\eta(x,y,t)$, since particles are free to move with fluid velocity,
the kinematic boundary condition is
\beq
{ w_{ \eta}}  = {\bf u_{ \eta}}  \cdot {\bf \bigtriangledown}\eta  
  + \frac{\partial \eta}{\partial t} 
\eeq 
and at bottom ${\it z= -h(x,y)} $  
\beq
 w_b = -u_b \cdot  {\bigtriangledown} h   
\eeq 
where $ {\bf u_{ \eta}}=(u_{ \eta},v_{ \eta}) $ is 
two component horizontal surface velocity.  $ \bigtriangledown \eta =(\eta_{ x},
\eta_{ y}),  $ 
 $ \bigtriangledown h =( h_{ x}, h_{ y} ) $ refer to 
horizontal derivative with respect to 
x and y in all subsequent calculations.   The horizontal 
component for vorticity field
$ \bf s =( s_{ y}, -s_{ x})$ can be described as,
\beq
\frac{\partial u}{\partial z} - \mu^{ 2} {\bigtriangledown} w ={\bf s}
\eeq  
with  ${\bf u} = (u,v)$ as two component horizontal field whereas
vertical component of vorticity expressed as
\beq
-s_{ z} = \frac{\partial u}{\partial y} - \frac {\partial
  v}{\partial x}
\eeq

This is straightforward calculation from  equation (6) and (8)
which is the beginning equation in three dimensional vorticity 
field formulation.  
\beq
\mu^{ 2} {\bigtriangledown}^{ 2}w + \frac{\partial^{ 2}w}
{\partial z^{ 2}} = - {\bigtriangledown} \cdot s = S_{ w}
\eeq
 $w$ represents the vertical 
velocity of the flow.

In the above equation, once $w$ solved, horizontal component of velocity
$  u,v $ can be solved from vorticity relation.
In weakly hydrostatic case ( $ 0 < \mu^{ 2} \ll 1 $ ), solution is
typically obtained from iterative perturbation procedure with
successive correction term up to $\mu^{ 2}$.
 
In case of breaking waves where vorticity is very strong, so $ ({\partial
u}/ {\partial z} \sim O(1)). $ We assume solution as, $ u = u_{ o}
+ \mu^{ 2} u_{ 1} + O(\mu^{ 4}) $ and $ w = w_{ 0} +
\mu^{ 2} w_{ 1} + O(\mu^{ 4}) $ for horizontal and vertical
velocity component.

Under this assumption, Poisson equation becomes 
\beq
     \frac{\partial^{ 2} w_{ 0}}{\partial z^{ 2}} = S_{ w}
\eeq
\beq
    \frac{\partial^{ 2}w_{ 1}} { \partial z^{ 2} } = - \left[ 
\frac{\partial^{ 2}w_{ 0}} { \partial x^{ 2} } +
\frac{\partial^{ 2}w_{ 0}} { \partial y^{ 2} }  \right]
\eeq
$ w_{ 0}$,  $ w_{ 1}$ can be calculated from bottom boundary conditions 
using equationn (7) separately where the boundary conditions are,  
\beq   
 w_{ {b0}} = - u_{ {b0}} \cdot  {\bigtriangledown}h 
\eeq  and 
\beq
  w_{ {b1}} +  u_{ {b1}}\cdot
 {\bigtriangledown}h = 0 
\eeq at bottom boundary {\it z = -h } 
 
Since at any other depth $ {\bf z} = z_{ r}$, w is constrained by
continuity equation only, so the equation follows
\beq
\frac {\partial w}{\partial z} {\rbrack_{ z}}_{ r}
 = - {\bf \bigtriangledown} \cdot u_{ m}  {\rbrack_{ z}}_{
 r} + \frac {\partial u}{\partial z} \cdot {\bf \bigtriangledown}
 z_{ r} {\rbrack_{ z}}_{ r} 
\eeq 
where $u_{ m} $ is velocity at any arbitrary depth $z_{ r}$ .
In Boussinesq type equation, one may take depth average or any
intermediate velocity for horizontal velocity between bottom and free
surface as reference velocity.  In the wave breaking zone where the
vorticity is developed non uniformly, the equations become simpler
with the choice of depth average velocity which includes contribution
from surface vorticity gradient.  We assume solution for velocity
comes also from rotational contribution due to vorticity at the wave
surface. So the velocity has both potential as well as rotational
component, $ u = u_{ p} + u_{ r} $ , $ w = w_{ p} + w_{ r}
$ We solve $w_{ 0}, w_{ 1} $ and $ u_{ 0}, u_{ 1} $ at any
depth $z_{ r}$
\beq
 \frac {\partial w_{ 0}} { \partial z } 
 { ]_{z}}_{ r} = - {\bf \bigtriangledown} \cdot (u_{m}
  - z_{r}  s {\rbrack_{ z}}_{ r}) +
z_{ r}   {\bf \bigtriangledown} \cdot s
\eeq  
\beq
\lbrack  \frac {\partial w_{ 1}} { \partial z } {\rbrack_{ z}}_{ r}=
\lbrack {\bf \bigtriangledown} w_{ 0} {\rbrack_{ z}}_{ r} 
\cdot {\bf \bigtriangledown} z_{ r} 
\eeq  
\beq
 \frac {\partial u_{ 0} } {\partial z} = s  
\eeq
and
\beq
 \frac {\partial u_{ 1}} {\partial z} =  {\bigtriangledown} w_{ 0}
\eeq 
with boundary condition 
 $ \lbrack u_{ 0} {\rbrack_{ z}}_{ r} = u_{ r} $ and $
\lbrack u_{ 1} {\rbrack_{ z}}_{ r} = 0 $ Equations (4) - (16)
form basic shallow water Boussinesq equations.

\section{ Equation for horizontal velocity }
 
In the surf zone, vorticity grows very strongly as a non uniform
function over depth.  Following Shen [2000], we define reference
velocity as $ \tilde u = \bar u + \bigtriangleup \bar u - \eta s_{
\eta} $ in terms of depth average velocity $ {\it \bar u} $ and
magnitude of vorticity at free surface $ s_{ \eta} $ with the
assumption of $ \bigtriangledown \cdot s \ne 0 $.  we set here 
$ z_{ r} = \eta $ as linear calibration for    $ z_{ r} 
= r(\eta + h) - h $ 
does not hold
here in presence of nonuniform velocity as wave dispersion properties
change both spatially and temporally with vorticity.  And boundary
condition can be set as
\beq
\frac{\partial w}{\partial z}|_{ \eta} = {\bf \bigtriangledown} \cdot \tilde u
+ \eta({\bf \bigtriangledown} \cdot s_{ \eta})
\eeq

Integrating equation (9) from bottom to surface and applying boundary
condition to (16) we get $w_{ 0}$ as ,
\beq
w_{ 0} = w_{ {b0}} - ( -  {\bf \bigtriangledown} \cdot 
\tilde u + \eta  {\bf \bigtriangledown} \cdot s_{ {\eta}} ) H_{ z} -
S_{ {w0}} 
\eeq
where 
$$  S_{ {w0}} = \int \int (-\bigtriangledown \cdot s )  dz  dz $$  
is the vertical velocity distribution generated by horizontal
divergence of vorticity added to surface velocity.
 
Now, once $w_{ 0}$ is calculated, $u_{ 1}$ can be calculated
from eqn (15) with surface boundary condition $ \lbrack u_{ 0}
\rbrack_{ \eta} = u_{ m} $ and $ \lbrack u_{ 1} \rbrack_{
\eta} = 0 $
 
Finally, we calculate horizontal velocity as 
\beqr
u(z)  &=& u_{ \eta} - \int_{ z}^{ \eta} s dz +
{\mu}^{ 2} ( S_{ {wl}} - \bar S_{ {wl}}) \nonumber \\
&+&
\frac{\mu^{ 2}}{2}({H_{ \eta}^{ 2} } - {H_{ z}}^{ 2} )  
{\bigtriangledown}(\bigtriangledown \cdot \tilde u - \eta\bigtriangledown 
\cdot s_{ \eta})  \nonumber \\ 
&+& {\mu}^{ 2} ( H_{ \eta} - H_{ z}) \left[ 
{\bf \bigtriangledown} ((\tilde u + \eta s_{ \eta})\cdot {\bf \bigtriangledown }h )
+ 
({\bf \bigtriangledown} \cdot \tilde u - \eta \bigtriangledown \cdot 
s_{ \eta})
\bigtriangledown h \right] \nonumber \\
&+&  O(\mu^{ 4})
\eeqr

which on averaging over depth yields, 
\beqr
\bar u  &=& u_{ \eta} - \bigtriangleup \bar u  
+ \frac{\mu^{ 2}}{3}
{H_{ \eta}}^{ 2} \bigtriangledown ( 
\bigtriangledown \cdot \tilde u - \eta 
\bigtriangledown \cdot s_{ \eta}) \nonumber  \\
&-&  \frac{\mu^{ 2}}{2} H_{ \eta}\lbrack 
\bigtriangledown  (\tilde u + \eta s_{ \eta})\cdot \bigtriangledown h
 - ( \bigtriangledown \cdot 
\tilde u - \eta  \bigtriangledown \cdot s_{ \eta})\bigtriangledown h \rbrack 
 + O(\mu^{ 4})
\eeqr  

$ \bigtriangleup \bar u = \frac{1}{H_{ \eta}} \int_{ {-h}}^{
\eta} \bigtriangleup u(z) dz $ is the average surface velocity
contribution due to vorticity and it is significant for suspended
sediment particles in the flow.  The term $ \bigtriangleup {\bf u(z)}
= \int_{ z}^{ \eta} s dz $ is the change due to depth variation
of vorticity $ {\bf S} $.  The total water depth $ H_{ z} $ and
surface elevation $ H_{ \eta} $ are taken as $H_{ z}$ = z + h
and $H_{ \eta} = \eta + h $

 The contribution for velocity has 
and rotational component apart from potential due to vorticity 
generation. After we redefine  $  H_{ \eta}=d $ and
 $ z={H_{ z}}/{H_{ \eta}} $, we express potential and rotational
component up to order $ O(\mu^{ 2}) $ as
\beqr
u_{ p}(z) &=& \bar u_{ p} +  \frac{\mu^{ 2}}{2} (\frac{1}
{3} -  z^{ 2}) {d^{ 2}} 
 \bigtriangledown( \bigtriangledown \cdot \bar u_{ p}) \nonumber \\
&+& {\mu^{ 2}} (\frac{1}{2}  -  z)d 
\left[ \bigtriangledown(\tilde u_{ p} \cdot 
\bigtriangledown h) + ( \bigtriangledown \cdot \tilde u_{ p})
\bigtriangledown h \right]
\eeqr
\beqr
u_{ r}(z) &=&  \tilde u_{ r} - \bigtriangleup u(z)  
+ \eta s_{ \eta} 
 + \mu^{ 2} 
( S_{ {wl}}  -  \bar S_{ {wl}}) \nonumber \\
&-& \frac{\mu^{ 2}}{2} (\frac{1}{3} - z^{ 2})d^{ 2} {\eta}
\bigtriangledown
 (\bigtriangledown \cdot s_{ \eta})  \nonumber \\
&-&  {\mu^{ 2}} 
(\frac{1}{2} - z)d {\eta} \lbrack  
(\bigtriangledown \cdot s_{ \eta})\bigtriangledown h -
\bigtriangledown(s_{ \eta} \cdot \bigtriangledown h) \rbrack 
\eeqr 
Similar expressions for vertical velocity are  
\beqr
w_{ p}(z) &=& - (h+z) \bigtriangledown \cdot u_{ p}(z) \nonumber \\  
&=& -  \bigtriangledown \cdot\lbrack (h+z) \tilde u_{ p} \rbrack
- \frac{\mu^{ 2}}{2}(\frac{1}{3}-z^{ 2}) \bigtriangledown \cdot \lbrack
 d^{ 2} (h + z) (\bigtriangledown (\bigtriangledown 
\cdot \tilde u_{ p}) \rbrack \nonumber \\
&-&   \mu^{ 2}(\frac{1}{2}-z)\lbrack \bigtriangledown
\cdot (h+z) ( 
\bigtriangledown (u_{ p}
\cdot \bigtriangledown h) + (\bigtriangledown \cdot u_{ p}) 
\bigtriangledown h) \rbrack 
\eeqr
\beqr
w_{ r}(z) &=& -\bigtriangledown \cdot (h+z) \tilde u_{ r} -
 \bigtriangledown \cdot \lbrack(h+z) \eta s_{ \eta} \rbrack \nonumber 
\\
&-&  \frac{\mu^{ 2}}{2} \lbrack  \bigtriangledown 
\cdot (h+z) 
(\frac{1}{3} - z^{ 2})d^{ 2}
{\bigtriangledown}(\eta \bigtriangledown \cdot s_{ \eta}) \rbrack 
\nonumber \\
&+& {\mu^{ 2}} \bigtriangledown \cdot \lbrack ( \frac{1}{2} -z)d
\lbrack \bigtriangledown(\eta s_{ \eta}.\bigtriangledown h) 
- ( \bigtriangledown \cdot \eta s_{ \eta})\bigtriangledown h \rbrack 
\eeqr

\section{ Breaking Model [fully nonlinear case] } 

Conventional time dependent Boussinesq equations for surface wave
height and consequent breaking term calculation are very straight
forward and published previously in case of irrotational waves.  Here
we take up fully nonlinear calculation as vorticity becomes a large
fraction of water depth in the surf zone or shoaling waves.  So, while
developing Boussinesq equations for horizontal momentum, we retain up
to order O($\delta^{ 2})$ and O($\delta\mu^{ 2})$ in our fully
nonlinear calculation. Fully nonlinear Boussinesq equations for long
wave have been derived by Mei [1983] for flat bottom and by Wei et
al [1995] for variable bottom surface in case of irrotational
wave.  Shen [2000] addressed problems in developing generalized
three dimensional irrotational propagating wave field to include
rotational motion in general did not describe the vorticity breaking
terms.  For horizontal propagation of waves, the three dimensional
problem can be reduced in terms of two horizontal velocity by
integrating over depth and retaining up to order $ O(\delta^{ 2}) $
and $ O(\delta\mu^{ 2})$ As horizontal velocity is governed by
momentum equation at the surface $\eta$ by,
\beq
 \frac {D u}{D t} |_{ \eta} = ( \frac{D w}{D t} 
|_{ \eta} + 1 )
{\bigtriangledown} \eta 
\eeq
In the surf zone region of sloping beach, waves break due to high
vorticity and the breaking of wave later being converted to
turbulence. So horizontal variation of water depth ${\it h(x,y)} $
must be considered in this case.  We express surface propagation
equation in terms of average velocity description and total time
derivative of horizontal momentum can be written as,
\beq
\frac { D \bar u}{Dt} | _{ \eta} 
  =  \frac{\partial u}{\partial t}|_{ \eta} + u_{ \eta}\cdot 
(\bigtriangledown u) |_{ \eta} 
\eeq
where surface velocity is given by, 
\begin{eqnarray}
u_{ \eta}  &=& \bar u + \eta S_{ \eta} -
\frac{\mu^{ 2}}{3}d^{ 2} \bigtriangledown(
\bigtriangledown \cdot \tilde u - {\eta}\bigtriangledown \cdot s_{ \eta})
\nonumber \\
&+& \frac{\mu^{ 2}}{2}d \lbrack   \bigtriangledown(\tilde u -
\bigtriangleup \bar u |_{ {-h}} + \eta s_{ \eta}) \cdot
\bigtriangledown h ) -
(\bigtriangledown \cdot \tilde u -\eta \bigtriangledown \cdot
s_{ \eta})\bigtriangledown h \rbrack 
\end{eqnarray}
We consider $ \bigtriangledown H_{ \eta} = \bigtriangledown \eta +
\bigtriangledown h $ for wavy bottom
\beqr
\frac{Du}{Dt}|_{ \eta} 
&=&  \frac{\partial \tilde u}{\partial t}  +
+ \eta \frac{\partial s_{ \eta}}{\partial t}
+ \tilde u \cdot \bigtriangledown \tilde u 
- \frac{\mu^{ 2}}{3}d^{ 2}\lbrack 
\bigtriangledown(\bigtriangledown
\cdot \frac{\partial \tilde u}{\partial t} - {\eta}{\bigtriangledown}
\cdot \frac{\partial s_{ \eta}}{\partial t}) \nonumber \\
&+& \tilde u \cdot \bigtriangledown 
 ( \bigtriangledown \cdot {\tilde u} 
-{\eta}\bigtriangledown \cdot s_{ \eta}) + 
\frac{\mu^{ 2}}{2}d \lbrack \bigtriangledown(\frac{\partial \tilde u}
{\partial t} + {\eta}\frac{\partial s_{ \eta}}{\partial t} )
\cdot \bigtriangledown h  \nonumber \\
&-& (\bigtriangledown \cdot \frac{\partial \tilde u}
{\partial t} 
- {\eta}\bigtriangledown \cdot \frac{\partial s_{ \eta}}
{\partial t})\bigtriangledown h \nonumber \\
&+& \tilde u \cdot \bigtriangledown \rbrack  + O(\mu^{ 4})
\end{eqnarray}

This long wave momentum equation upon simplification over flat bottom
case can be compared to the one derived by Shen [2000] The
vertical velocity can be obtained similarly, \\
\begin{eqnarray}
\frac {Dw}{Dt} |_{ \eta} &=& \frac{\partial w }
{\partial t} |_{ \eta} + 
u_{ \eta} \cdot \bigtriangledown w_{ \eta} + 
w\frac{\partial w}{\partial z}
|_{ \eta}    
\eeqr  
So, we can write the horizontal momentum equation as, 
\beqr
\frac{\partial \tilde u}{\partial t}  &+&  \tilde u \cdot \bigtriangledown
\tilde u  + \bigtriangledown \eta
= \frac{\mu^{ 2}}{3}d^{ 2} \{ \bigtriangledown(\bigtriangledown
\cdot \tilde u)\cdot \bigtriangledown 
 \tilde u + \bigtriangledown(\bigtriangledown \cdot 
\frac{\partial \tilde u}{\partial t})  \nonumber \\
&+&  (\tilde u 
\cdot \bigtriangledown(\bigtriangledown \cdot \tilde u)
\} - \mu^{ 2 }d \{ \bigtriangledown \cdot \frac{\partial \tilde u}
{\partial t} -d^{ 2} \bigtriangledown(\bigtriangledown \cdot \frac
{\partial \tilde u}{\partial t}) + (\bigtriangledown 
\cdot u_{ \eta})^{ 2} \nonumber \\
&-& \tilde u \cdot \bigtriangledown (\bigtriangledown \cdot \tilde u) \} 
 \bigtriangledown {\eta}   
\eeqr
 $ \tilde u $ is defined in previous section.  In contrast to the
 result by Shen [2000], additional contribution factor here
 arises from vorticity variation which is significant for surf zone
 wave.  Wei et al [1995] also breaking term for irrotational long
 wave momentum equation over a variable bottom wave. The intermediate
 depth velocity $ z_{ \alpha} $ is being used there proportional to
 h instead of depth average velocity used here which may not be valid
 inside the fluid.  The use of $ z_{ r} $ in our approach avoids
 this difficulty. Finally we try to generalize equation by solving
 vorticity from vorticity transport equation in next section.

\section{ Vorticity transport equation in breaking zone} 
Madsen and Svendsen [1983] used a cubic vertical distribution
of rotational velocity based on roller jump data which can not
considered in three dimension case as it is not guaranteed to bring
accuracy in the simulation.  So we try to solve vorticity function
from Reynold stress based equation.
\beq
\frac{\partial u}{\partial t} + ( u \cdot  
\bigtriangledown) u = -\frac{1}{\rho}{ \bigtriangledown p}
\eeq
Taking the curl on both sides and use vorticity function $ s =
\bigtriangledown \times u $ we get,
\beq
\frac{\partial s}{\partial t} -(s \cdot  
\bigtriangledown)s
+ ( u \cdot \bigtriangledown) s = \nu 
{\bigtriangledown}^{ 2} s
\eeq
$(s \cdot \bigtriangledown) s $ is "vorticity stretching" factor due
to change in gradient in vorticity. This term leads to change of
rotation of material particles present in the flow to the
beach. Contribution of this term can not be incorporated from two
dimension roller jump data.
 
We generalize the equation in three dimension as
\beqr
\frac{\partial s} {\partial t}  
&+& {\delta u} 
\frac{\partial s}{\partial x} + {\delta v}\frac{\partial s}{\partial y} 
+ {\delta w}\frac{\partial s}{\partial z} 
- {\delta s}\frac{\partial u}{\partial x} - {\delta s}\frac{\partial v}
{\partial y} -{\delta s}\frac{\partial w}{\partial z} \nonumber \\
&=& \nu \lbrack {\mu}^{ 2}\frac{\partial^{ 2} s}{{\partial x}^{ 2}}
+ {\mu}^{ 2}\frac{\partial^{ 2} s}{{\partial y}^{ 2}}
+ \frac{\partial^{ 2} s}{{\partial z}^{ 2}} \rbrack
\eeqr 
After changing the variable from (x,y,z,t) to wave following
coordinates $ (x,y,\sigma,t)$, we write the vorticity equation as
\beqr
\frac{\partial s}{\partial t} &-& \frac{\delta \sigma}{(h+\delta\eta)}\frac
{\partial \eta}{\partial t} \frac{\partial s}{\partial \sigma} + 
\delta u (\bigtriangledown \cdot s) - 
{\delta}s (\bigtriangledown \cdot u) 
- \frac{\delta }{(h+\delta \eta)}\lbrack s\frac{\partial w}
{\partial \sigma} - w \frac {\partial s}
{\partial \sigma} \rbrack  \nonumber  \\
&-& {\delta}^{ 2} \frac{\sigma u }
{(h+\delta \eta)} (\bigtriangledown  \cdot
\eta)  \frac{\partial s}{\partial \sigma} + {\delta}^{ 2}
\frac{\sigma s}{(h+\delta\eta)}(\bigtriangledown \cdot \eta) 
\frac{\partial u}{\partial \sigma} \nonumber \\
&=& \nu \lbrack {\mu^{ 2}} \bigtriangledown^{ 2}s +
 \frac{1}{(h+\delta\eta)}
\frac{\partial^{ 2}s}{\partial \sigma^{ 2}} \rbrack +O(\mu^{ 2})
+ O(h_{ x}) + O(h_{ y})
\eeqr
The boundary conditions in new coordinate system are, 
\begin{subequations}
\begin{align}
s(\sigma=1,t) &= s(x,y,t) ;
s(\sigma=0,t) &= 0 ; 
s(\sigma,t=0) &= 0 
\end{align}
\end{subequations}

After we redefine $ s = \Omega + \sigma {\omega}_{ s} $, which
transforms the equation to which is easier to solve:
\begin{eqnarray}
\frac{\partial \Omega}{\partial t} &+&
 {\sigma}\frac{\partial \omega_{ s}}{\partial t}
 - {\delta}\frac{\sigma}{(h +{\delta}{\eta})}\lbrack
\sigma \frac{\partial \eta}{\partial t} 
- \frac{\partial \eta}
{\partial t}\frac{\partial \Omega}{\partial \sigma}\rbrack \nonumber \\
&+& {\delta}u(\bigtriangledown \cdot \Omega) - \frac{\delta \sigma}
{(h+ \delta \eta)} (\bigtriangledown \cdot \eta) \frac{\partial \Omega}
{\partial \sigma} - \frac{\delta \sigma^{ 2}}{(h+\delta \eta)}
(\bigtriangledown \cdot \eta)  \nonumber \\
&+& \frac{\delta u}{(h+\delta \eta)} \frac{\partial \Omega}
{\partial \sigma} + \frac{\delta w}{(h+\delta \eta)} \omega_{ s} 
 \nonumber \\
&-& \frac{\delta \Omega}{(h+ \delta \eta)} 
\frac{\partial w }{\partial \sigma} - \frac{\delta \sigma 
\omega_{ s}}{(h+ \delta \eta)} \frac{\partial w}{\partial \sigma} 
\nonumber \\
&-&  \delta \Omega (\bigtriangledown \cdot u) - {\delta}{\sigma}
\omega_{ s} (\bigtriangledown \cdot u) + \frac{ \delta^{ 2} \sigma
\Omega}{(h + \delta \eta)}(\bigtriangledown \cdot \eta) \frac{\partial u}
{\partial \sigma} + \frac{\delta^{ 2} \sigma \Omega}{(h+\delta \eta)}
(\bigtriangledown \cdot \Omega)\frac{\partial u}{\partial \sigma} \nonumber \\
&+& \frac{ \delta^{ 2} \sigma \omega_{ s}}{(h+\delta \eta)}
(\bigtriangledown \cdot \eta) \frac{\partial u}{\partial \sigma} 
\end{eqnarray} 
with new boundary, 
\beq
\Omega(\sigma=1, t) = 0 
\Omega(\sigma=0, t) = 0   
\eeq
with initial condition $ \Omega(\sigma,t=0) = 0 $. This additional
equation can be solved numerically as done by Briganti et al
[2004] for the two dimensional case or an analytical solution can
be formulated as shown by Veeramony \& Svendsen [2000].  The
analytical solution can be calculated by assuming $\Omega =
\omega^{ {(1)}} + {\delta}\omega^{{(2)}} $ which gives first and
second solution as
\underline {O(1) Problem}
\beq
\frac{\partial \omega^{ {(1)}}}{\partial t} + {\sigma}
\frac{\partial \omega_{ s}}{\partial t} = 
\frac{\nu}{h^{ 2}}\frac{\partial^{ 2}\omega^{ {(1)}}}
{\partial t^{ 2}}
\eeq
where the solution is 
\beq
 F_{ n}^{ {(1)}} = (-1)^{ n} \frac{2}{n\pi} 
\frac{\partial \omega_{ s}}{\partial t} 
\eeq
assuming $ - \sigma \frac{\partial \omega_{ s}}
{\partial t} = {\sum_{n=1}}^{\infty} F_{n}^{1} 
\sin n\pi\sigma   $  
And to solve $ \omega^{ {(1)}} $, assume 
$ \omega^{ {(1)}} = \Sigma_{ n} 
G_{ n}^{ {(1)}} sin n \pi \sigma $ 
which gives  zeroth order solution as 
\beq
G_{ n}^{ {(1)}} = (-1)^{ n} \frac{2}{n\pi} \int_{ 0}^{ t} 
\frac{\partial \omega_{ s}}{\partial \tau} 
e^{ n^{ 2}\pi^{ 2}
\kappa(\tau-t)} d \tau
\eeq
To consider \underline { O($\delta$)Problem } 
\beq
\frac{\partial \omega^{ (2)}}{\partial \sigma}  
- \frac{\nu}{h} \frac{\partial^{ 2} \omega^{ (2)}}
{\partial \sigma^{ 2}}  = F^{ (2)} 
\eeq
where 
\beqr
F^{ (2) } &=& 
\frac{\sigma}{h}\frac{\partial \eta}
{\partial t} \frac{\partial \omega^{(1)}}{\partial \sigma}
 - \frac{\sigma^{ 2}}{h}\frac{\partial \eta}{\partial t} 
-\frac{\sigma}{h}(\bigtriangledown \cdot \eta)\frac{\partial 
\omega^{ (1)}}{\partial \sigma}  - \frac{\sigma^{ 2}}
{h}(\bigtriangledown \cdot \eta)  + u( \bigtriangledown \cdot 
\omega^{ (1)}) \\ \nonumber
&+& \frac{u}{h} \frac{\partial 
\omega^{ (1)} }{\partial \sigma} 
\eeqr
 To solve above equation, assume
$ \omega^{(2)} = \Sigma_{ n}^{ {(2)}} 
sin n\pi \sigma $ 
where solution becomes
\beq
G_{ n}^{ {(2)}} = 2\int_{ 0}^{ 1} 
F_{ n}^{ {(2)}}e^{ n^{ 2}\pi^{ 2}\kappa(\tau -t)}
d \tau
\eeq
with
\beq
F_{ n}^{ {(2)}} = 2\int_{ 0}^{ 1} 
F^{(2)}sin n\pi\sigma
d \sigma  
\eeq 
The solution for vorticity s becomes,
\beq
s = \sigma \omega_{ s} + \Sigma_{ 1} {G_{ n}}^{ (1)}
sin n {\pi}{\sigma} + \Sigma_{ 1} {G_{ n}}^{ (2)} 
sin n {\pi}{\sigma} 
\eeq
To solve breaking term, we need value of $ \omega_{ s} $ for
boundary and eddy viscosity value as input data.

\section{Conclusion}

Finally we conclude here by developing a most generalized form of
fully nonlinear Boussinesq equations for wave propagation in surf zone
region with variable bathymetry with vorticity distribution from
Vorticity Transport Equation(VTE). 
In this wave breaking zone, vorticity generated by the shear stress of
current is very strong, so contribution to the surface velocity due to
vorticity variation has significant contribution in fluid flow.
These extra terms in generalized equation complicate the numerical
technique as these terms are present in the equation in multiple form
of equations for vorticity components which has to be solved in
coupled solution technique. Veeramony [2000] used simplied the
formulation by taking constant eddy viscosity value but this
oversimplified case may bring inaccuracy in calculation. Briganti
et.al [2004] formulated a numerical technique scheme to solve
VTE  using generalized depth variable eddy
viscosity  $ \nu= \nu(x,y) $ in two dimension case. In three
dimensional formulation, the nonlinear terms in the vorticity
transport equation(VTE) will complicate the calculation and so proper
numerical technique have to be developed. This work is under way.

\section{Acknowledgment}
This work was supported by Office of Naval Research Lab (NRL) under grant
 [GR001820].

\end{document}